# HIGH GRADIENT SUPERCONDUCTING CAVITY WITH LOW SURFACE EM FIELDS AND WELL-SUPPRESSED HOMS FOR THE ILC

N. Juntong and R.M. Jones
The Cockcroft Institute, Daresbury, Warrington, Cheshire WA4 4AD, UK.
The University of Manchester, Oxford Road, Manchester, M13 9PL, UK.

*Abstract*

We present an optimized geometry for a 1.3 GHz superconducting cavity in which the surface electromagnetic fields have been minimized and the bandwidth of the fundamental mode has been maximized. We refer to this design as the New Low Surface Field (NLSF) cavity [1]. Earlier work focused on properties of the fundamental mode. Here we additionally study higher order modes (HOMs), means of damping them, and short range wake fields. A two-band circuit model is employed in order to facilitate rapid characterization of cavity HOMs.

## INTRODUCTION

The International Linear Collider (ILC) is a lepton machine designed to produce collisions between electrons and positrons at a centre of mass energy of 500 GeV and with the potential to be upgraded to 1 TeV [2]. This machine is based on superconducting (SC) linear accelerators. The present baseline design prescribes a working accelerating gradient of 31.5 MV/m. Increasing the accelerating gradient has the beneficial effect of potentially reducing the overall footprint of the accelerator. However, a larger accelerating gradient leads to increased surface electric and magnetic fields. An enhanced surface electric field ($E_s$) often gives rise to field emission of electrons and can lead to both electron capture in the field of the rf accelerating cavities and to rf breakdown. On the other hand, increasing the surface magnetic field ($B_s$) can quench the superconducting properties of the cavity. It is critically important to minimize both the surface magnetic field and the surface electric field with respect to the accelerating field ($E_a$). Another figure of merit is the fractional operating bandwidth $k_c \approx (\omega_\pi - \omega_0)/\omega_{\pi/2}$ of the monopole mode. The importance of this is made clear by considering the separation of the next-nearest mode to the accelerating mode for these 9-cell cavities:

$$\Delta\bar{\omega} = (\omega_\pi - \omega_{8\pi/9})/\omega_{\pi/2} = (\pi/18)^2 k_c \quad (1)$$

In order to prevent excitation of spurious modes, fabrication tolerances must be carefully controlled. Thus, we focused our design study on modifying the cavity shape with a view to optimize three quantities: $E_s/E_a$, $B_s/E_a$ and $k_c$; the former two are minimized, whilst the latter is maximized.

This paper is organized such that the procedure used to obtain the optimized parameters is described in the next section. The section following thereafter analyses the rf properties of modes that will be excited in these NLSF cavities. The penultimate section presents a two-band circuit model and an analysis of the short range wakefields. Some concluding remarks are presented in the final section.

## OPTIMAL CELL SHAPE

The geometrical shape of a SC cell is illustrated in Fig. 1. The NLSF cavity is based on a vertical wall ($\theta = 0$). An additional alternative design with a non-vertical wall ($\theta \neq 0$) is also studied. In all cases, the cell length is fixed at half the operating wavelength ($L = \lambda/2 = 115.304$ mm). We assign the iris thickness (*a*) to that of the LSF design [3] and the equator radius ($R_{eq}$) to that of the LL [4, 5] design. Two parameters are varied in order to obtain an optimal rf design: *b* and *B*. The goal is to minimize $E_s/E_a$, $B_s/E_a$ and maximize $k_c$. For this purpose we performed detailed simulations with the code Poisson Superfish [6]. These have been subsequently validated using the finite element code HFSS [7]. Fig. 2 shows a select series of optimization results in comparison with present designs.

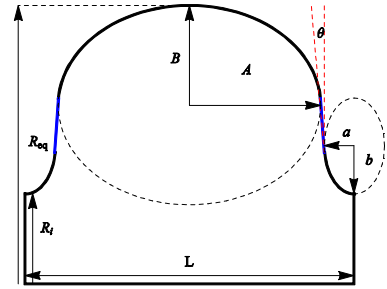

Figure 1: Parameterisation for a single SC cell

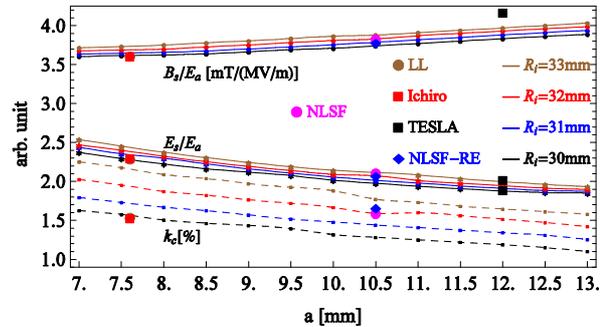

Figure 2: Variation of rf figures of merit with iris thickness and radius, compared to ongoing designs.

The NLSF shape has a 8.5% lower $B_s/E_a$ field compared to TESLA, and $E_s/E_a$ is 13% lower than the LL design.

Both $E_s/E_a$ and $B_s/E_a$ are comparable to the LSF design; however the bandwidth is superior as it is ~26.5% wider. We also investigated re-entrant shapes ($\theta<0$), at a fixed iris radius of 32mm, again focusing on optimizing three figures of merit [8]. This results in an optimal angle of -2°.

After these optimisations are complete, the end cells must also be designed to ensure the accelerating field is ~99% flat. The HOM couplers are then carefully matched into the accelerating cavity. The HOM pipe radius in the NLSF design was chosen to be identical to that in the LL cavity. Table 1 summarizes the results of an optimisation study for NLSF and NLSF-RE. The NLSF shape has inferior parameters compared to the re-entrant shape. Despite this, we nonetheless opt for this shape for a full design study as in fabricating the cavity surface cleaning is considerably simpler for this shape.

Table 1: Optimal NLSF and NLSF-RE cavities

| Parameters | NLSF | | NLSF-RE | |
|---|---|---|---|---|
| | Mid cell | End cell | Mid cell | End cell |
| $R_i$ [mm] | 32.0 | 38.0 | 32.0 | 32.0 |
| $R_{eq}$ [mm] | 98.58 | 98.58 | 98.58 | 98.58 |
| A [mm] | 47.152 | 48.152 | 49.0 | 50.0 |
| B [mm] | 31.35 | 30.5 | 35.30 | 35.0 |
| $a$ [mm] | 10.5 | 9.5 | 10.5 | 10.0 |
| $b$ [mm] | 15.5 | 12.5 | 17 | 15.0 |
| $k_c$ [%] | 1.59 | | 1.66 | |
| $E_s/E_a$ | 2.11 | | 2.07 | |
| $B_s/E_a$ [mT/(MV/m)] | 3.83 | | 3.78 | |
| R/Q 9 cells [$\Omega$] | 1127 | | 1134 | |

## DESIGN OF POWER COUPLER

Various shapes for the rf power couplers for NLSF cavity have been investigated and are illustrated in Fig. 3. In all cases we aim at achieving an external quality factor $Q_{ext} = 3.5 \times 10^6$, similar to the ILC. This ensures that the perturbation on the accelerating field flatness is minimised. In all simulations, the eigenmode module of CST MWS [9] was utilised to obtain $Q_{ext}$ and these results are displayed in Fig. 3. Out of several possible couplers, we opt for the standard TTF-III coupler [10] for our NLSF cavity. In order to obtain the requisite Q, the coupler is placed 45 mm away from cavity entrance and with an antenna penetration depth of 6 mm.

## CIRCUIT MODEL ANALYSIS

We employ a two-band circuit model [11], previously applied to the TESLA cavity [12], to our NLSF cavity. This model allows the lower dipole bands to be well represented. However, the discrepancy becomes larger when applied to higher bands. In order to overcome this limitation we add an additional term to the circuit model, which includes the effect of an iris finite thickness. This model is enhanced with an additional parameter $\chi$, which requires a further eigenmode simulation in order to determine this parameter. The circuit model for the dipole modes results in:

$$\left(\frac{1-\eta\cos\phi}{\omega_r^2}-\frac{1}{\omega^2}\right)\left(\frac{1-\hat{\eta}\cos\phi}{\hat{\omega}_r^2}-\frac{1}{\omega^2}\right)-\chi\frac{\bar{\eta}^2\sin^2\phi}{\omega_r^2\hat{\omega}_r^2}=0 \quad (2)$$

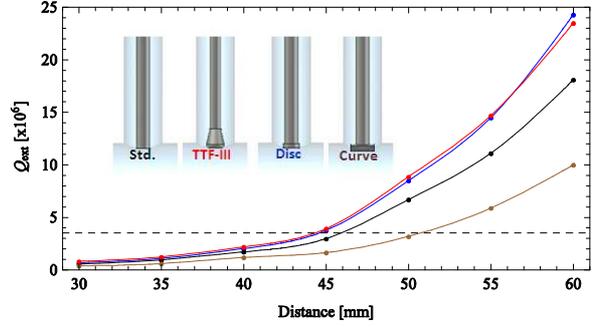

Figure 3: $Q_{ext}$ of various couplers vs distance away from cavity entrance. Each individual antenna penetrates 6mm into the cavity and is shown inset (black, red, blue, brown, indicate the standard coaxial without a tip, TTF-III type, disc tip, and curved tip, respectively).

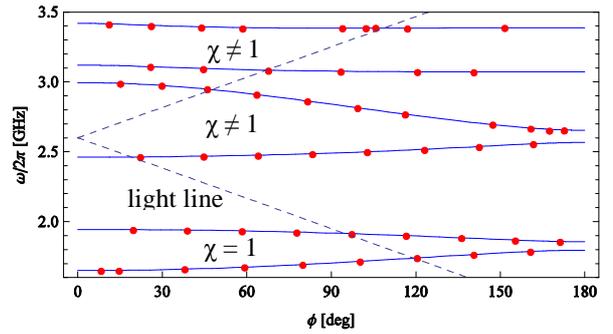

Figure 4: Modified two-band circuit model up to the 6th dipole band, indicated by solid lines, versus eigenfrequencies obtained from HFSS, indicated by dots.

Here the quantities $\omega_r/2\pi$, $\hat{\omega}_r/2\pi$ correspond to the resonant frequencies of the TM and TE components of the cells, respectively. Also, $\eta, \hat{\eta}$ are the usual coupling constants of the TM and TE bands, respectively and $\bar{\eta}=\sqrt{\eta\hat{\eta}}$ is the cross-coupling between TM and TE components in the thin iris approximation. The phase advance per cell is represented by $\phi$ and $\omega/2\pi$ is the mode frequency. Figs. 4 and 5 show the results from this modified two-band circuit model of NLSF cavity. This model allows the both the dispersion curves and the R/Q values to be well-predicted up to the 6th dipole band and hence the required wakefield suppression can be designed accordingly. On the other hand, the short range wakefield is essentially prescribed according to the average iris radius and this must be kept to within tolerable limits in order to minimise emittance dilution. In the next section we investigate the short range wakefield.

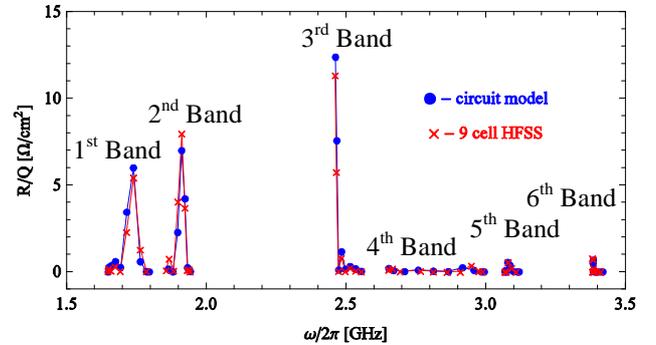

Figure 5: R/Q from a circuit model applied to NLFS vs HFSS results up to 6th dipole band.

## SHORT RANGE WAKEFIELD

We perform detailed numerical simulations on both the longitudinal and transverse wakefield of our NLSF cavity, with the code ECHO2D [13]. Here we investigate the periodic nature of the wakefield by undertaking simulations with a string of 8 cavities. In addition, we compared this to an analytical model based on the periodic behaviour of a pill box cavity [14]. The analytical form of the longitudinal and transverse short range wake functions are given by:

$$w_l(s) = -\theta(s)\left[Ae^{-\sqrt{s/s_0}} + B\frac{\cos(\omega s^\alpha)}{\sqrt{s} + Cs^\beta}\right] \quad (3a)$$

$$w_t(s) = \theta(s)\left[D\left(1-\left(1+\sqrt{s/s_1}\right)e^{-\sqrt{s/s_1}}\right) + E\sqrt{s}\right] \quad (3b)$$

In order to obtain these wake functions, we vary the associated parameters at a prescribed $\sigma$ and fit them to ECHO2D simulations. For example, to obtain the longitudinal wake potential, the wake function is convolved with a Gaussian, with $\sigma$=300μm, and A, B, C, $\omega$, $s_0$, $\alpha$, $\beta$ are varied in order to fit to ECHO2D simulations. A similar procedure is followed for the transverse wake function. We characterised the NLSF cavity with the following parameters A=36.5, B=0.35, C=320, $\omega$=1030, $\alpha$=$\beta$=1 and $s_0$=2.44x10$^{-3}$ for the longitudinal wake function and D=90, E=555 and $s_1$=9.5x10$^{-4}$ for the transverse wake function. We then varied $\sigma$ and compared the analytical prediction with the results of ECHO2D simulations. These results are displayed in Fig. 6 and Fig

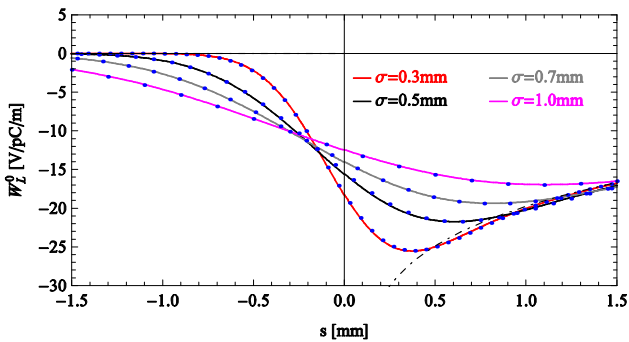

Figure 6: Analytical longitudinal wake function (black dot-dashed line), analytical wake potentials (coloured solid lines) for the NLSF structure and numerical wake potentials (blue dots) for Gaussian bunches with σ = 0.03, 0.05, 0.07 and 0.1 mm.

7 for the longitudinal and transverse potentials, respectively. It is evident that, for a wide range of σ, the analytical model provides an excellent prediction for the wake potential.

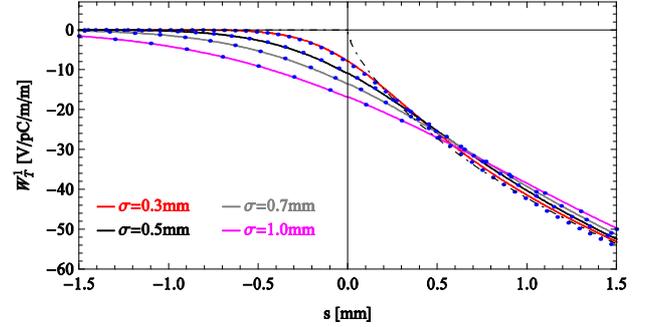

Figure 7: Analytical transverse wake function (black dot-dashed line), analytical wake potentials (coloured solid lines) for the NLSF structure and numerical wake potentials (blue dots) for Gaussian bunches with σ = 0.03, 0.05, 0.07 and 0.1 mm.

## CONCLUDING REMARKS

A design for the NLSF cavity has been presented, together with an alternative NLSF-RE design. A fundamental power coupler, based on the ILC coupler, has been designed for our NLSF cavity to obtain the requisite coupling. Short-range wakefields have been calculated and analytical models employed to aid the characterisation. Future work will be focussed on the HOM coupler and on the long-range wakefield.

## ACKNOWLEDGEMENT

We have benefited from discussions at the MEW group meetings at the Cockcroft Institute. N.J. receives support from the Royal Thai Government and the Thai Synchrotron Light Research Institute.

## REFERENCES

[1] N. Juntong, R.M. Jones, SRF2009, THPPO024, 2009
[2] The International Linear Collider Reference Design Report, 2007.
[3] Z. Li, C. Adolphsen, LINAC08, THP038, 2008.
[4] J. Sekutowicz et al., JLAB TN-02-023, 2002.
[5] J. Sekutowicz et al., PAC05, TPPT056, 2005.
[6] K. Halbach, R.F. Holsinger, Part. Accel. **7**, 213-222, 1976.
[7] www.ansoft.com/products/hf/hfss/
[8] R.L. Geng, Physica **C 441**, 145-150, 2006.
[9] www.cst.com/Content/Products/MWS/
[10] W.-D. Möller for the TESLA collaboration, 9th SRF Workshop, Santa Fe, Vol. 2, p. 577-581, 1999
[11] K.L.F. Bane, R.L. Gluckstern, Part. Accel. **42**: p. 123-169, 1993.
[12] N. Juntong et al., EPAC08, WEPP085, 2008.
[13] I. Zagorodnov et al., J. Comp. Physics. **191**:525, 2003.
[14] K.L.F. Bane, SLAC-PUB-9663, 2003.